\documentclass[preprint,showpacs,preprintnumbers,amsmath,amssymb,nohyperref]{revtex4}

\usepackage[most]{tcolorbox}
\usepackage{makecell}
\usepackage{booktabs}
\usepackage{tabularx}
\usepackage{mathtools}
\usepackage{dsfont}
\usepackage{mathrsfs}
\usepackage{wrapfig}
\usepackage{xurl}
\usepackage[T1]{fontenc}
\usepackage{pgfplots}

\usepackage{lastpage}

\usepackage{graphicx}
\usepackage{mathrsfs}
\usepackage{bm}
\usepackage{color,graphicx}
\usepackage{graphicx,colordvi}
\usepackage[ngerman,english]{babel}
\usepackage{caption}
\usepackage{subcaption}
\usepackage{setspace}

\newcommand{\bel}[1]{\begin{equation}\label{#1}}

\def\be{\begin{equation}}
\def\ee{\end{equation}}
\def\bea{\begin{eqnarray}}
\def\eea{\end{eqnarray}}
\def\l{\label}

\def\ms{\medskip}
\def\siml{\;\hbox{\kern.1em \lower.7ex \hbox{$\sim$} \kern-1.12em
\raise.5ex \hbox{$<$} \kern.1em}}
\def\simg{\;\hbox{\kern.1em \lower.7ex \hbox{$\sim$} \kern-1.12em
\raise.5ex \hbox{$>$} \kern.1em}}

\def\d{\hbox{d}}

\def\l{\label}

\def\d{\hbox{d}}

\def\siml{\hbox{\kern.1em \lower.6ex \hbox{$\sim$} \kern-1.12em
          \raise.6ex \hbox{$<$} \kern.1em }}
\def\simg{\hbox{\kern.1em \lower.6ex \hbox{$\sim$} \kern-1.12em
          \raise.6ex \hbox{$>$} \kern.1em }}

\ms


\begin{document}

\pdfoutput=1

\title{CHEMICAL POTENTIAL OF A HADRONIC FIREBALL IN THE FREEZE-OUT STAGE}

\author{Yaroslav D.~Krivenko-Emetov}
\email{krivemet@ukr.net, y.kryvenko-emetov@kpi.ua}
\affiliation{National Technical University of Ukraine, 03056, Kiev}

\author{Andriy I.~Smetana}
\affiliation{National Technical University of Ukraine, 03056, Kiev}

\begin{abstract}
This article explores the van der Waals gas model proposed to describe the hadronic stages of nuclear fireball evolution during the cooling stage. Two different models were proposed for the early and late stages of hadronization. At the initial stage, a two-component meson model consisting of $\pi^0$ and $\pi^+$ mesons was suggested, and at the later stage, a two-component nucleon model consisting of protons and neutrons was proposed. The interaction potential for both models was represented by a rectangular well, and the statistical sum was calculated using the saddle-point method. The analytic expressions for pressure and chemical potentials obtained from the model were compared with the corresponding numerical results of other authors obtained earlier using quantum chromodynamics (QCD) methods. The possibility of applying and using the effective chemical potential is also analyzed.%
\end{abstract}

\maketitle
\section*{Introduction}
Experiments that observe an elliptical flow in non-central collisions of heavy nuclei at high energies indicate that a state of quark-gluon plasma appears and thermalization occurs. This is due to the fact that particles collide with each other more than once. The substance in this state can be characterized by the thermodynamic quantities of temperature, viscosity, density, and others. After the quark-gluon plasma cools, a hadron gas is formed, which can be described in terms of statistical models of hadron gas, such as the van der Waals (vdW) model \cite{stachel}-\cite{yeng}. The vdW model is especially useful as it takes into account the repulsion effect that prevents high density at high temperatures. The Grand Canonical Ensemble (GCE) is a suitable mathematical formalism for the phenomena observed in heavy-ion collisions as the number of particles is not fixed. The vdW model proposed in \cite{gorenstein} introduces the phenomenological parameters of the radii of the hard-core $R_{ii}$ and $R_{ij}$ that significantly change the yield of particles with different types $N_i$.

Various authors have proposed the development of the vdW model to describe more subtle effects in the dependence of the hadronic gas pressure on density (e.g.  \cite{krivemet}, \cite{gorvov} ). In a multicomponent gas, the parameter $a$ corresponding to attraction transforms into parameters $a_{ij}$, and the repulsive parameter $b$ transforms into parameters $b_{ij}$. The effective potential parameters depend on the effective radii of repulsion $R_i^0$ and attraction $R_i$. However, the vdW model cannot be properly developed when considering a finite nuclear system. For nuclear collisions, a nuclear fireball with dimensions $<r>~\sim 7-100$ Fm is observed.  In this case, the GCE formalism leads to the use of a double sum, which can be transformed into a double integral, which can be integrated by the saddle point method.%

The problem has been solved for a two-component system using the GCE formalism, leading to the use of a double sum that can be transformed into a multidimensional integral. A model presented in \cite{krivemet1} was believed to be applicable for collisions of heavy nuclei at CERN, with the assumption that characteristic temperatures do not exceed the temperatures at which new particles can be generated. The temperatures of the nuclear fireball are around $T < 130$ MeV, and the model should have a transparent nonrelativistic limit while considering the law of conservation of the total number of nucleons without the generation of new particles.%
\begin{figure}
\begin{center}
\includegraphics[scale=0.27]{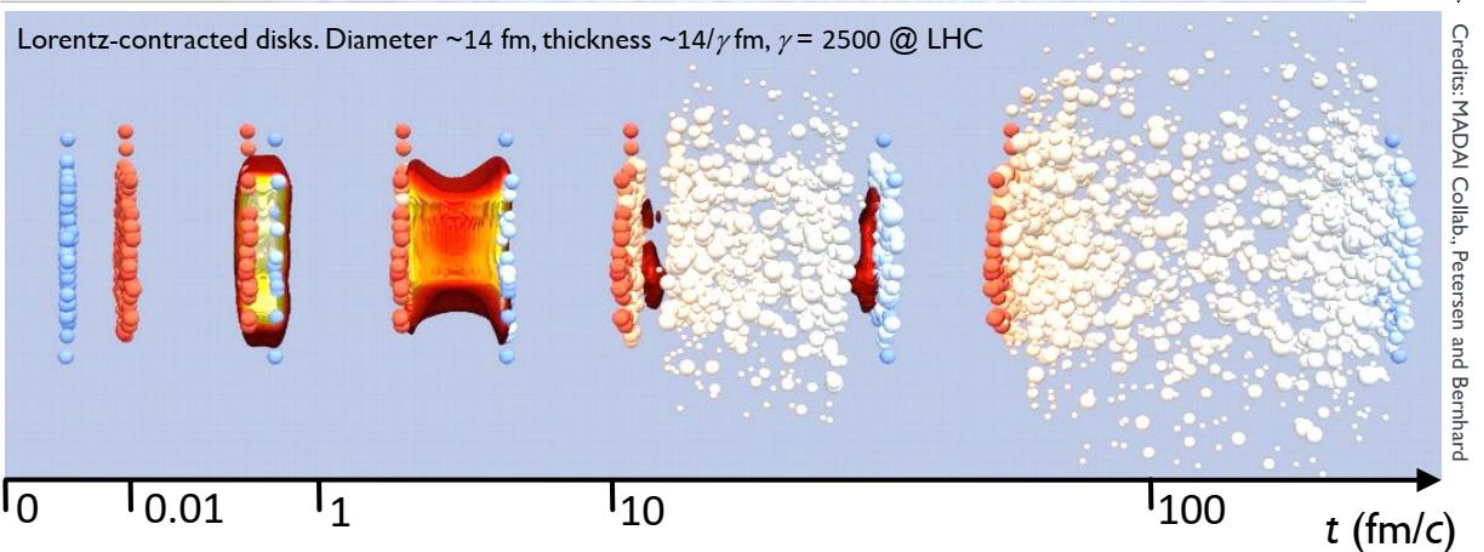}
\caption{The successive stages of the evolution of a nuclear fireball  \cite{freeze}}
%\figcaption{The successive stages of the evolution of a nuclear fireball  \cite{freeze}}
\label{pic1}
\end{center}
\end{figure}

%\begin{figure}
%\begin{center}
%\centering
%\includegraphics[scale=0.27]{QGP3}
%\caption{The successive stages of the evolution of a nuclear fireball  \cite{freeze}}
%\label{pic1}
%\end{center}
%\end{figure}

In Fig. \ref{pic1}, the different stages of a nuclear fireball's evolution are presented, including the initial state of two touching ultrarelativistic nuclei, followed by a hot and superdense nuclear system, the quark-gluon phase, hadronization and chemical freeze-out, and finally, kinetic freeze-out. A more detailed and comprehensive explanation of the mathematical model proposed in \cite{krivemet1} is provided in the article, which includes an evaluation of some finer effects such as additional corrections for pressure, density, and root-mean-square (RMS) fluctuations. For temperatures above the production threshold, a new two-component meson model \cite{krivemet2},\cite{krivemet3} is proposed, where the number of mesons is not conserved when $T > 135$ MeV.%

In conclusion, the study of the quark-gluon plasma is an active field of research that involves many theoretical and experimental efforts. The understanding of its properties is essential in understanding the early universe, the behavior of matter under extreme conditions, and the dynamics of heavy-ion collisions.
In the work  \cite{fukush}, based on quantum chromodynamics (QCD) calculations, it was obtained that if one approaches the critical point along the first-order phase boundary, the corresponding shape of the potential looks like that of a second-order phase transition around a non-zero order parameter, as shown in Fig. \ref{pic2}.:
\begin{figure}
\begin{center}
\includegraphics[scale=1]{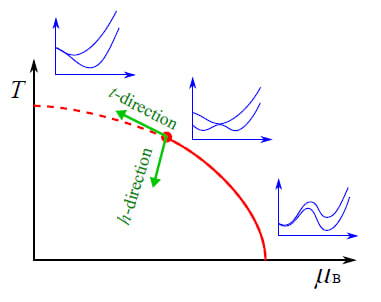}
%\figcaption{Schematic representation of the chemical potential and the forms of the equations of state in the crossover, critical region and first order region, respectively, from left to right}
\caption{Schematic representation of the chemical potential and the forms of the equations of state in the crossover, critical region and first order region, respectively, from left to right}
\label{pic2}
\end{center}
\end{figure}

\section{One-component vdW gas}
\label{sec-vdw}

Based on Fig. \ref{pic1}, the lifetime of a nuclear fireball ($ t>10^{-21} c$) is estimated to be much longer than the characteristic nuclear interaction time of $t' \sim 10^{-23}-10^{-24}$ c. The relaxation time of the system is estimated to be of the order of $\tau \sim 10^{-20}-10^{-22}$ c for small local fireball volumes. As a result, it can be assumed that a local statistical equilibrium is established in the subsystem at each moment of time exceeding the relaxation time. This implies that the local fireball region is quasi-stationary, and statistical physics methods can be used to describe it.

Since all thermodynamic potentials, entropy, and volume are extensive, the potentials of the entire system can be defined as the sum of the corresponding thermodynamic potentials of quasi-closed subsystems. This approach is based on the assumption that at each moment of time, a standard representation of the partition function of a rarefied quasi-ideal van der Waals gas in the canonical ensemble can be provided for such quasi-closed subsystems. This quantity has a specific form in the approximation of pair interaction and when the condition $B(T)N/V \ll 1$ is satisfied \cite{landau}.

\be\l{eq1}
Z(V,T,N)  = \frac{1}{N!} \phi(T,m)^N(V-B(T)N)^N,
\ee
where, respectively, $N$  and  $m$    are the number and mass of particles, $V$   and $T$  are the volume and temperature of the gas.
Formula ($\ref{eq1}$)  uses the notation \cite{gorenstein}($\hbar=c=k_B=1$):
%\begin{fleqn}
$$\phi (m,T)=$$
%\end{fleqn}

\be\l{eq:14444}
%\begin{equation}
%\begin{flushleft}
=\frac{1}{2 \pi^2} \int_0^\infty p^2 \exp{\left(-\frac{\sqrt{m^2+p^2}}{T}\right)} dp=\frac{m^2 T}{2\pi^2}K_2(m/T),
\ee
%\end{equation}
where  $K_2(z)$ is the modified Bessel function, and the second virial coefficient in  $\eqref{eq1}$ has the form:
\be\l{eq:1555}
B(T)= \frac{1}{2}\int_0^\infty (1-\exp(-U/T)) dV
\ee
and includes pairwise interaction of particles, $U =  U_{ij}$, $(i \neq j)$.

In relativistic limit $ m \gg T $   one can easy obtain, given the asymptotes of the Bessel function:
$$
\phi (m,T) \sim\left(\frac{m T}{2\pi}\right)^{3/2}\exp{\left(-\frac{m}{T}\right)}.
$$
This formula further leads to the effect of exponential suppression of the particle yields with large mass, which is important in the study of quark-gluon plasma.

The pressure in the system is easy to find from the partition function:
\be\l{eq2}
\mathcal{P}(V,T,N)= T\frac{\partial}{\partial V}
%16 ln%
\ln[Z(V,T,N)]  =  \frac{TN}{V-B(T)N} .
\ee
Note that if the Stirling formula is used in the partition function for the factorial:$$N! \approx \sqrt{2 \pi N}(N/e)^N,$$ then the final pressure formula (\ref{eq2}) will not change.

The model \cite{krivemet2},\cite{krivemet3} being used for calculations of subsystems involves applying methods of statistical physics, assuming local statistical equilibrium and the fulfillment of the statistical boundary condition of $N \rightarrow N_A$, where $N_A$ is the Avogadro constant. This assumption is justified at the initial stages of evolution, as the number of particles generated in a fireball is around 3-5 thousand during high-energy nucleus-nucleus interactions. However, at later stages of evolution, the assumption becomes doubtful, as the number of nucleons in the nonrelativistic limit is limited by the baryon number conservation law and is equal to $N\sim200$. Nonetheless, the practical application of the van der Waals equation often goes beyond the conditions under which the virial approximation has been obtained. Therefore, the approximation is believed to be sufficiently justified, especially as it is always possible to restrict calculations to the first stage. Although the saddle point method is used when $B(T)<0$, the final formulas can be extended to a region where the second virial coefficient $B(T)$ is not necessarily negative.

From the partition function  $Z(V,T,N)$  one can also get: free energy $F(V,T,N)=-T\ln[Z(V,T,N)]$, chemical potential
$$\mu=\left(\frac{\partial F(V,T,N)}{\partial N}\right)=$$
%\begin{scriptsize}
\be\l{eq7771}
=T \left[ \ln (N/V)- \ln (\phi(T,m)) + \frac{ 2B(T) N}{V}\right]
\ee
%\end{scriptsize}

and the derivative of the chemical potential which in the statistical limit has the form:
$$(  \partial \mu /\partial N ) =-(\partial P/\partial V ) (V/N)^2=$$
\be\l{eq877791}
=  \lim_{N \to N_A} \left( \frac{T}{N}+\frac{ 2B(T)T}{V}\right) \rightarrow \frac{ 2B(T)T}{V}.
\ee
Then, we obtain the Grand partition function (GPF)  $\mathcal{Z}(V,T,\mu)$  from the partition function  $Z(V,T,N)$  taking into account the above physical considerations (see, e.g. \cite{kubo},\cite{feynman}):

\be\l{eq3}
\mathcal{Z}(V,T,\mu)= \sum_{N} \exp\left(\frac{\mu N}{T}\right)Z(V,T,N).
\ee
At high temperatures (which, for example, are realized during collisions of heavy ions in the GCE, and  $ \bigtriangleup N/T \rightarrow  dN'$ ) one can turn from the sum to the integral using the Euler-Maclaurin formula. In this case, the first integral term remains and the logarithm of the statistical sum is introduced into the exponent. Let's denote this indicator by $\Phi(N')$ :
\\
$\mathcal{Z}(V,T,\mu) = T\int_{0}^{ \infty} \d N'
\exp\left(\mu N'+ \ln[Z(V,N'(T))]\right)=$
\be\l{eq8}
=T\int_{0}^{ \infty} \d N'
\exp\left( \Phi(N')\right).
\ee
\begin{figure}
\begin{center}
\includegraphics[scale=0.32]{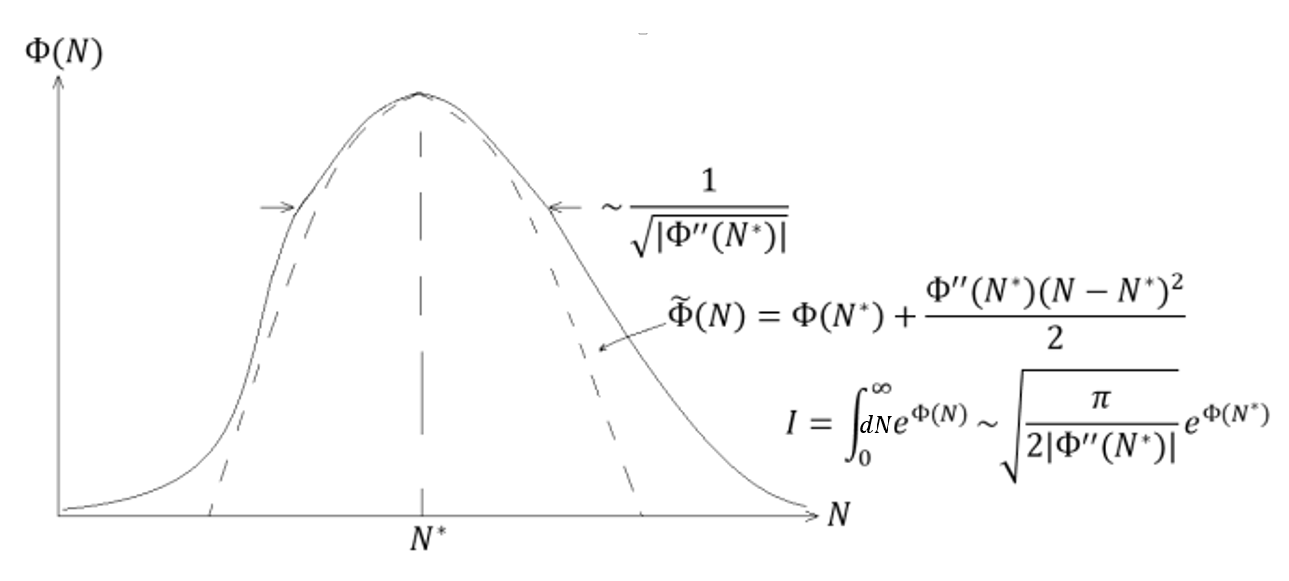}
\caption{Graphic explanation of the saddle point method}
\label{pic3}
\end{center}
\end{figure}

The saddle point method( see Fig.\ref{pic3}) is utilized for further integration, as the integrand has a clear maximum at high temperatures. The maximum point ($N^{\ast}$) for the integrand is obtained in \cite{krivemet2},\cite{krivemet3} through the extremum condition imposed on the saddle point:
\begin{scriptsize}
\be\l{eq5}
\mu^*(N^*)=- \left(T\frac{\partial}{\partial N} \ln [Z(V,T,N)]\right)_{N=N^{\ast}} -N^*(\partial \mu /\partial N )_{N=N^*}\approx
\ee
\end{scriptsize}
\be\l{eq51}
\approx T \left[ \ln (N^*/V)- \ln (\phi(T,m) )\right] ,
\ee
where $\mu^*$   is the chemical potential at the saddle point.

As a result, obtaind \cite{krivemet2},\cite{krivemet3}:
$$
P(T,\mu^*)  \approx T\xi [1 - B(T)\xi -\ln {(  \partial^2 \Phi^* /\partial N^2 )_{N=N^* }}/(2V\xi)]
$$
\be\l{eq888}
  \approx T\xi[1-B(T)\xi -\ln {[B(T)\xi ]}/(2V \xi)],
\ee
where the saddle point,  $\xi=N^*(V, T, \mu^*)/V$, is defined according to (\ref{eq51}) and (\ref{eq7771}) as  $\xi =\phi (m,T) \exp{(\mu^*(\xi) /T)}$.
The parameter   $\xi$ can be eliminated from Eq.\ (\ref{eq888}) using the definition of density, which in the thermodynamic limit turns into the well-known formula \cite{landau}:
\begin{small}
\be\l{eq1000}
n= \frac{\partial P (T, \mu)}{\partial \mu} = \xi [1-2B(T)\xi] - \frac{1}{2V}  \rightarrow \xi [1-2B(T)\xi] .
\ee
\end{small}
In the thermodynamic limit ($N \rightarrow N_A$,  $V \rightarrow \infty$ ) the chemical potential of the saddle point $\mu^*$   from  Eq.\ (\ref{eq51}) when  $N^*=N/(1-2B(T)N/V)$  turns into the chemical potential $\mu$ ( $\mu^*   \rightarrow \mu$ ), which is determined by the well-known
 thermodynamic equation Eq.\ (\ref{eq7771}).

Both equations  (Eq. (\ref{eq888}) and Eq. (\ref{eq1000})) in parametric form (the saddle point  $\xi$  acts as a parameter) determine the relationship between pressure $P$ , temperature $T$ , and density $n$ . In \cite{krivemet2},\cite{krivemet3} was obtained the state equation in GCE by excluding explicitly this parameter from the system of Eq. (\ref{eq888}) and Eq. (\ref{eq1000}):
\be\l{eq7777}
P(T,n)  \approx  T n[1 + B(T)n] +dP.
\ee
Of course, the resulting state equation is implicitly a parametric equation, since the saddle point  $\xi$ (and, hence, $n$  ) determines the chemical potential $\mu$ according to Eq.\ (\ref{eq7771}) and Eq.\ (\ref{eq51}), as:
\be\l{eq7671}
n=\phi(T,m)\exp(\mu/T -2B(T)n)
\ee

The formula derived takes into consideration the pressure contribution from the finite volume of the system, denoted as $V_s$. The author admits that the nature of this contribution is not fully understood. It is possible that it is a non-physical outcome that can be reduced by considering additional terms of the expansion by the saddle-point method. However, until further analysis is conducted to confirm this, the author treats this contribution as real and proceeds to quantitatively evaluate it. It is worth noting that this contribution disappears in the thermodynamic limit, where there is no distinction between CE and GCE.

As we consider large but finite volumes, only the last term on $dP$ remains in \ref{eq7777}, which is similar to what was observed in the meson and nuclear fireball model discussed in Sec. 3 and Sec. 4:

\begin{large}
$dP=\lim_{V \to V_{s}} \frac{T}{2V} (1 + B(T)n -ln [B(T)n] )  \rightarrow$
\end{large}
\be\l{eq777}
 dP=-\frac{T ln [B(T)n]}{2V_s}.
\ee
If we disregard the correction obtained from the volume of dP and assume that $B(T)n<<1$, then by making the following substitution in the right-hand side of Eq. (\ref{eq7777}): $(1+B(T)n) \sim \exp(B(T)n)$, taking into account Eq. (\ref{eq7671}), it will become the following:
$$P(T,\mu)  \approx  T \phi(T,m)\exp(\mu/T - B(T)n) =$$
\be\l{eq2777}
= T \phi(T,m)\exp(\mu^{int}/T)=P^{id}(T, \mu^{int}).
\ee
Thus, the equation of state with interaction can be obtained by making the substitution $\mu->\mu^{int}=\mu-TB(T)n$ in the equation of state of the ideal gas.
These equations are density functionals, which, according to (\ref{eq7771}), at a fixed chemical potential, are found from the solution of a transcendental equation $n=\phi(T,m)\exp(\mu/T -2B(T)n)$.
Assuming $B(T)n<<1$, this formula can be replaced with (\ref{eq1000}) where, according to (\ref{eq51}), $\xi$ is expressed in terms of $\mu$:

\be\l{eq1015}
n \approx  \phi(T,m)\exp(\mu/T)[1-2B(T)\phi(T,m)\exp(\mu/T)] .
\ee

%Цей доданок зникає у термодинамічній межі ($ V_s \rightarrow\infty$), а рівняння яке залишається для частинок нульових радіусів притягання та відштовхування  $B(T) = 0$  переходить у рівняння ідеального газу:
%\be\l{eq8887777}
%P(T,\mu,n)= P^{id}\left( T, \mu \right) =T n^{id} ( T, \mu).
%\ee
The RMS fluctuations of pressure and density calculated by known formulas (see, e.g.,\cite{landau} \cite{fed}) give estimates of the found corrections to the corresponding quantities:
\be\l{eq190011}
<(\bigtriangleup P)> \sim T \sqrt{n/V}[1+B(T)n],
\ee
\be\l{eq100011}
<(\bigtriangleup n)> \sim \frac{1}{\sqrt{nV^3}}[1-B(T)n].
\ee

\section{Multicomponent vdW gas}
\label{mult-vdw}
%\section{Two-component vdW gas}
%\label{sec-vdw2}
It is easy to extend the results obtained in Sec. 1 to the case of a two-component vdW gas ($i=1,2$)\cite{krivemet2}:

$\mu^*_i\rightarrow \mu_i=\left(\frac{\partial F(V,T,N_i,N_j)}{\partial N_i}\right),$

where $F(V,T,N_1,N_2)=-T \ln [Z(V,T,N_1,N_2)] $ is the definition of free energy a two-component vdW gas, whis density of components:
\be\label{eq339}
n_i= \partial P (T, \mu_i,\mu_j) /\partial \mu_i \sim \xi_i[1  -(2\xi_i B_{ii}+\xi_j (\tilde{B}_{ij}+\tilde{B}_{ji}))].
\ee
The virial expansion can be rewritten, taking into account  Eq. (\ref{eq339}), as a two-component vdW equation in the approximation  $b_{ij}N_i/V \ll 1$  and  $a_{ij}/Tb_{ij} \ll 1$) :
$$P(T,\mu_1,\mu_2)   = \frac{Tn_1}{1-b_{11}n_1-\tilde{b}_{21}n_2}+\frac{Tn_2}{1-b_{22}n_2-\tilde{b}_{12}n_1}$$
\bea\label{eq:33}
 -  n_1 (a_{11} n_1+\tilde{a}_{21} n_2)-n_2 (a_{22} n_2+\tilde{a}_{12} n_1)+dP,
\eea
where $dP$ takes into account the "finite size of the fireball":
$dP\cong -T n\frac{ ln [C(T,n_{11},n_{22})]}{2V}$, $C(n,T)=|n_{11}B_{11}n_{22}B_{22}-n_{12}\tilde{B}_{12}n_{21}\tilde{B}_{21}|$.
When formula (\ref{eq:33}) was derived, the expression   $
 \tilde{B}_{ij} \approx  \tilde{b}_{ij}- \tilde{a}_{ij}/T$ was used (see, e.g., \cite{krivemet}), and for each type of particles the corresponding parameters of attraction and repulsion were introduced:  $a_{ij}$, $\tilde{a}_{ij}  \approx 2 \gamma a_{ij} a_{ii}/(a_{ii}+a_{jj}) $,   $b_{ij}$, $\tilde{b}_{ij}=2 \frac{b_{ii}b_{ij}}{b_{ii}+b_{jj}}$,  $\gamma$ is a phenomenological parameter reflecting the complexity of the problem.

The above analysis can be extended to a multi-component VdW gas consisting of any number of different particles. By integrating over the particles' momenta and making some changes analogous to those made in the first example, an expression for the statistical sum of the multi-component (K-component) VdW gas can be obtained. Then, by integrating the obtained formula over the number of particles and making the corresponding changes to the formula, as was done in the case of a single-component gas, the corresponding expression for pressure can be obtained in the mathematical formalism of the Grand Canonical Ensemble\cite{krivemet3}:
$$P(T,\mu_1,...,\mu_K)=\sum \limits_{p=1}^{K}\left[ \frac{Tn_{p}}{1-\sum \limits_{(p\neq q)=1}^{K}(b_{pp}n_p+\tilde{b}_{pq}n_q)} \right]$$
\bea\label{eq:335}
 - \sum \limits_{(p\neq q)=1}^{K}n_p (a_{pp} n_p+\tilde{a}_{qp} n_q)+dP(T,\mu_1,...,\mu_K),
\eea
where $\mu_p=\left(\frac{\partial F(V,T,N_1,...,N_K)}{\partial N_p}\right) $ are the chemical potentials $(p = 1, ...,K)$.
%The functions $\xi_q$ satisfy the set of coupled transcendental equations where $A_p=$.
The particle densities $n_p= \partial P (T, \mu_1,...,\mu_K) /\partial \mu_p$ along with the pressure are obtained as the solutions of the system of related equations depending on the parameter of the saddle points $\xi_p$ $(p = 1, ...,K)$.

Due to inelastic reactions between hadrons in the HG model in the grand canonical ensemble formulation, there are no fixed numbers for $N_1, ..., N_K$. However, the conserved charges of baryonic number $B$, strangeness $S$, and electric charge $Q$ have fixed values. $B$ corresponds to the number of participating nucleons in the reaction, $S$ is zero, and $Q=eZ$ is 0.5eA for intermediate nuclei and 0.4eA for heavy nuclei. Using the grand canonical formulation is more advantageous at high temperatures, where the system properties are determined by the pressure function \ref{eq:335}. The chemical potentials $\mu_i$ (where $i = 1, ..., K$) are defined as a combination of the baryonic $\mu_B$, strange $\mu_S$, and electric $\mu_Q$ chemical potentials, with coefficients of expansion $(\gamma_{B})_i$, $(\gamma_{S})_i$, and $(\gamma_{Q})_i$ respectively.

In publications \cite{krivemet1} - \cite{krivemet3}, a two-stage model was proposed to describe the nuclear fireball. At the first stage, since the considered nuclear-nuclear collisions ($A+A$) have very high energies, more than 1 GeV per nucleon, the number of produced mesons is much larger than nucleons and it is suggested that the main contribution to the process is made mainly by mesons (meson stage of the fireball dynamics). Later, at the second stage, since the lifetime of mesons is much shorter than the lifetime of nucleons, it is proposed that the main contribution at the final stage is made by nucleons (nucleon stage of the fireball).

\begin{figure}
\begin{center}
\includegraphics[scale=1.1]{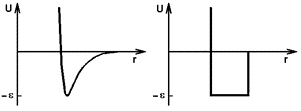}
\caption{Scalar part of the effective meson-meson potential}
\label{pic4}
\end{center}
\end{figure}

The meson model of the fireball is based on the following assumptions:
The collision between nuclei (A+A) generates high energies of over 1 GeV per nucleon. During the initial freeze-out stage, mesons dominate. To explain these interactions above the new particles' production threshold (T>135 MeV), the vdW model is generalized to a medium-sized meson fireball. The mean semiaxes of the ellipsoid and mass number of nuclei left in the fireball after the collision, <a>, <b>, and <A>, respectively, are used to estimate the average potential energy. The fireball is composed mostly of mesons, considering that the number of nucleons is significantly less than the number of mesons. Only two types of particles are considered, namely the $\pi^0$ and $\pi^+$ mesons, with the average internucleon energy not exceeding the production threshold of the heavy mesons. The $\pi^+$ meson production is twice as likely as the $\pi^0$ meson production(also, the lifetime of $\pi^+$ mesons is greater than the lifetime of $\pi^0$-mesons); thus, the densities are assumed to be equal to $k n_+=n_0$, where $n_0$ and $n_+$ are the densities of $\pi^0$ and $\pi^+$ mesons, respectively, and $k<1$. The potential energy of meson interaction is effective, and its scalar part is shown in Fig. \ref{pic4}.

It is assumed that the $\pi^0$ meson hard-core radius is much smaller than the $\pi^+$ meson hard-core radius. The effective radius of the $\pi^+$ meson is 0.46Fm, the $\pi^0$ meson is 0.01Fm, and the average volume of the meson fireball is estimated to be 600-1000 $Fm^3$

The nucleon model of the fireball is based on the following assumptions:
Mesons with a short average lifetime ($\tau \sim 10^{-8}-10^{-16} c$) dominate in the initial stages of freeze-out, leading to their quick decay. Baryons like protons and neutrons become dominant during the final stages of freezing. The finite volume size effects become evident at low density values, which corresponds to the last stages of fireball evolution. Although there is some uncertainty about the fireball's existence during these late stages, a generalization of the vdW model to the nucleon fireball was proposed in \cite{krivemet1} to describe nucleus-nucleus interactions during the last stage of freeze-out when new particles are not produced ($T<135$ MeV). To simplify the model, the average energies of internucleon collisions are restricted to not exceed the production threshold of other hadrons, and only two varieties (protons and neutrons) are considered. The density of protons and neutrons is assumed to follow from the conservation of baryon number, and the nucleon composition of colliding nuclei is assumed to be known. The effective potential of the interactions between protons and neutrons, protons and protons, and neutrons and neutrons can be represented using the same model as in Fig. \ref{pic4}. The hard-core radius of the proton is assumed to be known, while the radius of the neutron is much smaller than that of the proton.

\begin{figure}
\begin{center}
\includegraphics[scale=0.3]{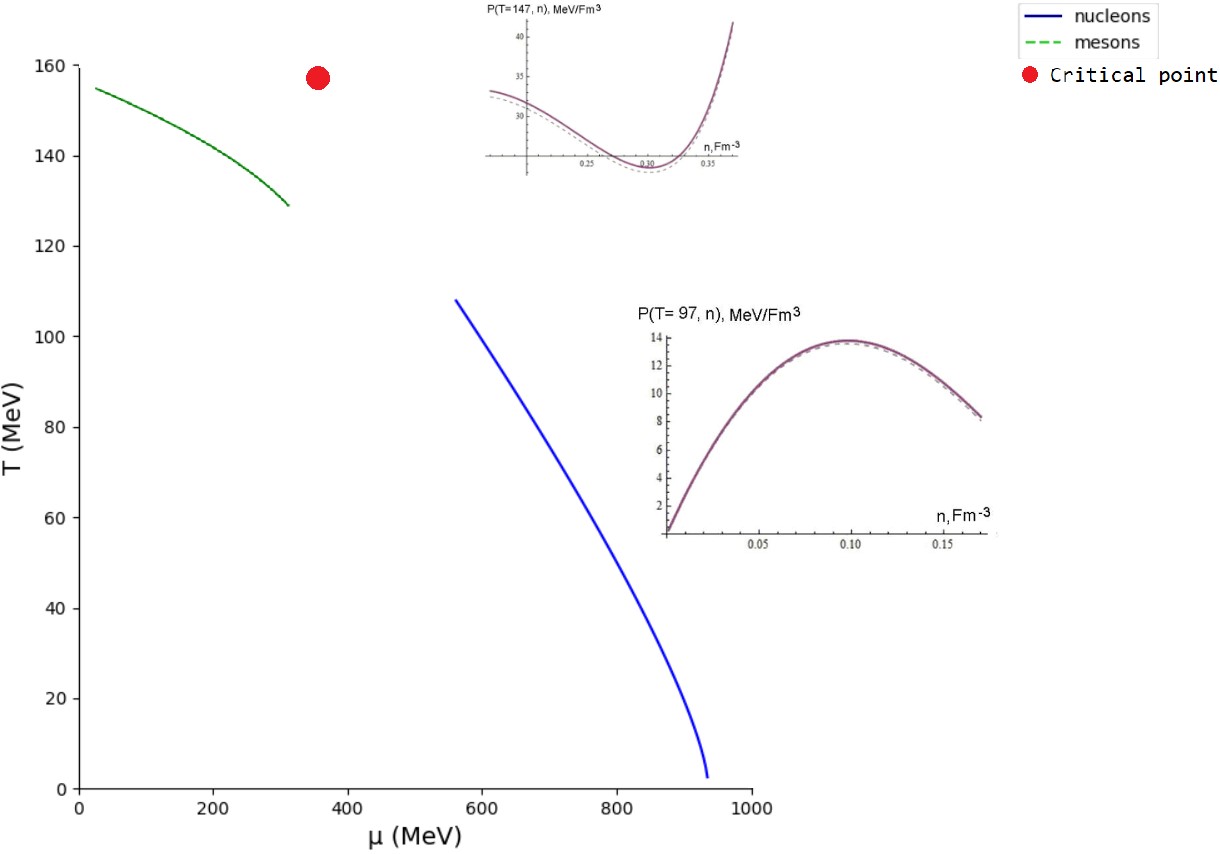}
\caption{The result of our calculations using formula \ref{eq7771} for the meson and nucleon stages of the evolution of the hadron fireball}
\label{pic5}
\end{center}
\end{figure}

\begin{figure}
\begin{center}
\includegraphics[scale=0.7]{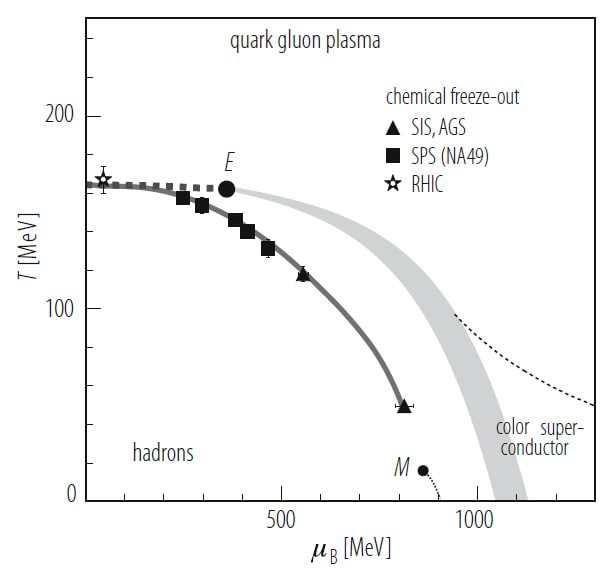}
\caption{Sketch of the QCD matter phase diagram in the plane of temperature T and baryo-chemical potential $\mu_B$(this figure is taken from \cite{stock1}). The parton-hadron phase transition line from lattice QCD \cite{Karsch1}–\cite{Karsch2}  ends in a critical point E. A cross-over transition occurs at smaller $\mu_B$. Also shown are the points of hadro-chemical freeze-out from the grand canonical statistical model}
\label{pic6}
\end{center}
\end{figure}

Thanks to the known relationship between the number of protons and neutrons in heavy nuclei, $n_p = k n_n$, where $k<1$, the two-component nucleon model essentially reduces to a one-component model. Similarly, since the lifetime of neutral pion mesons is much shorter than that of charged mesons, the same can be said for the two-component meson model. Interestingly, despite the crudeness of such a one-component approximation for the real multi-component vdW gas of the hadron fireball, as shown in Fig. \ref{pic5}, a good qualitative and quantitative agreement with the results of calculations by other authors is obtained for the chemical potential (see, for example, Fig. \ref{pic2} and Fig. \ref{pic6}).

\section{Approximation of an ideal gas with an effective chemical potential}
\label{approx}
In Sec. 1. (see \ref{eq2777}), it was shown that the Van der Waals gas pressure formula can be approximately rewritten as the pressure of an ideal gas with an effective potential $\mu^{int}=\mu-TB(T)n$ that takes into account interactions. However, this effective potential differs by $3TB(T)n$ from the chemical potential obtained from the known formula \ref{eq7771}.

The question arises as to how justified this approximation of the effective chemical potential is?

\begin{figure}
\begin{center}
\includegraphics[scale=0.45]{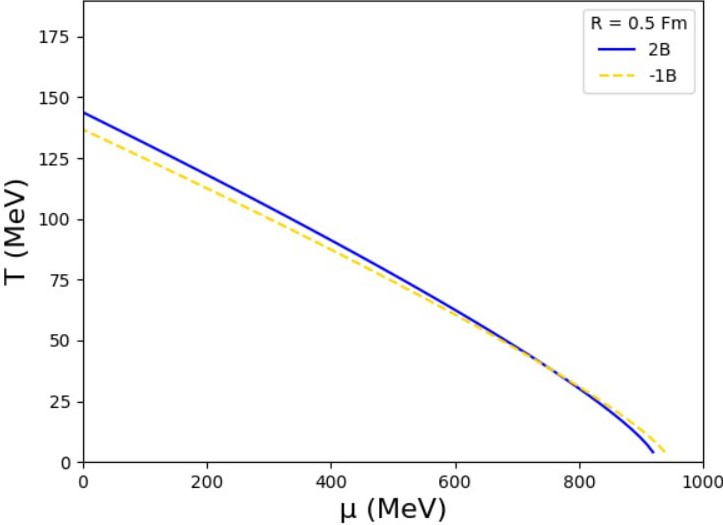}
\caption{Chemical potential calculated in the nucleon model of the fireball at hard core radii of $R^0=0.5$ Fm}
\label{pic7}
\end{center}
\end{figure}

\begin{figure}
\begin{center}
\includegraphics[scale=0.45]{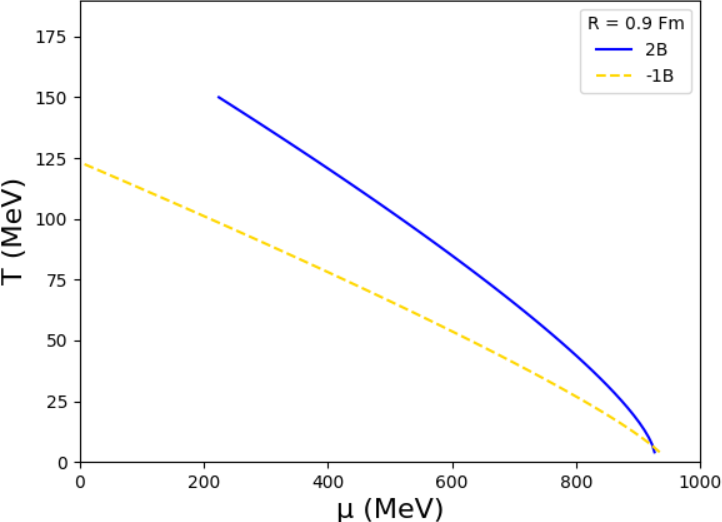}
\caption{Chemical potential calculated in the nucleon model of the fireball at hard core radii of $R^0=0.9$ Fm}
\label{pic8}
\end{center}
\end{figure}

\begin{figure}
\begin{center}
\includegraphics[scale=0.45]{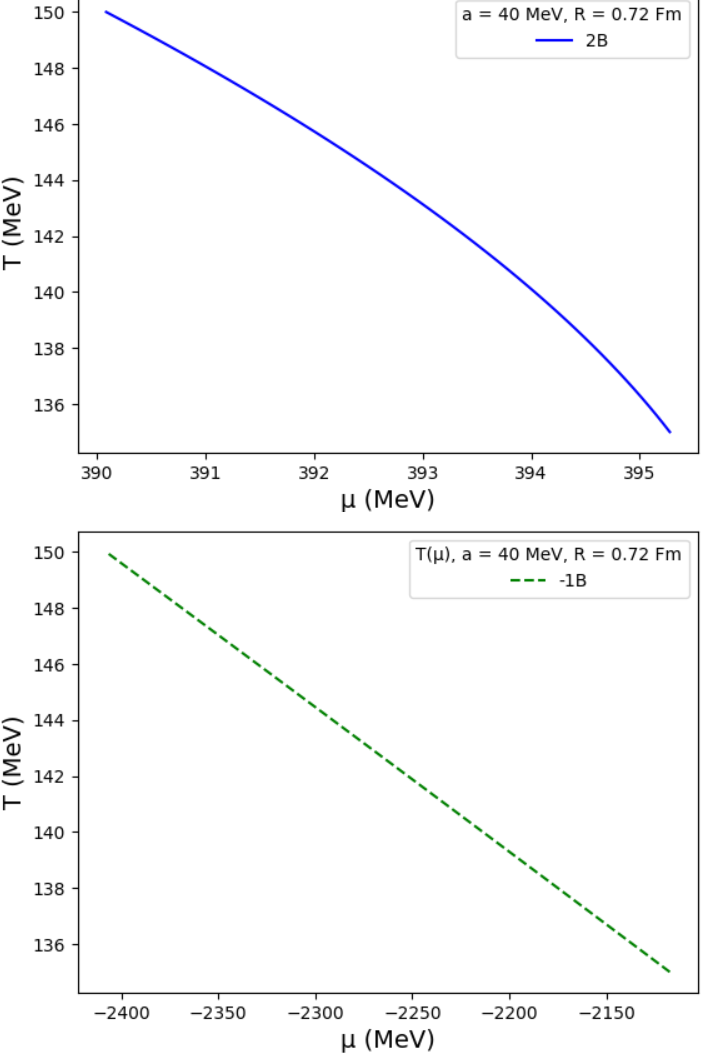}
\caption{Chemical potential calculated in the meson model of the fireball at hard core radii of $R^0=0.69$ Fm}
\label{pic9}
\end{center}
\end{figure}

Figures \ref{pic7}-\ref{pic9} show a comparison of the effective chemical potential and the real chemical potential at different parameter values for the nucleon and meson models.

It can be seen that while the nucleon model of the fireball at realistic radii of the solid nucleon crust is practically indistinguishable when replacing the real chemical potential with the effective one, for the meson model of the fireball at realistic parameter values, the effective chemical potential yields an explicitly unphysical result.

\section{Summary}
The paper provides a detailed description of obtaining the equation of state with corrections that take into account the finite size of the hadronic fireball, as well as the mean square fluctuations of pressure and density. The pressure correction disappears in the thermodynamic limit when, according to statistical physics, there is no difference between different statistical ensembles.

It is shown how, in the process of deriving the equation of state using the integration method near the saddle point, expressions for the chemical potentials are naturally obtained, which essentially serve as the extremum condition of the integrand.

The obtained formulas for chemical potentials, pressure, and density, obtained using the saddle point method, are used to analyze numerical data obtained by other authors using the lattice QCD calculations and critical parameters in nucleus-nucleus collisions at high energy.

The influence of interaction on the two-component gas is analyzed: (i) $\pi^0$ and $\pi^+$ mesons; (ii) baryons. The particles interact with a hard core potential at short distances and an attractive potential at long distances (effective radii of attraction).

As an example of the use of the obtained formulas at temperatures above the particle production threshold (T>135 MeV), a generalized van der Waals model was proposed for the asymmetric case of the two-component model ($\pi^0$ and $\pi^+$ mesons) with realistic parameters of the hard core and attraction.

At lower temperatures (T<90 MeV), a baryonic model was used, which takes into account the conservation law of baryonic charge.

It was found that despite the roughness of the one-component effective models for the equations of state and chemical potentials, they qualitatively, and sometimes quantitatively, agree with the results obtained using the lattice QCD calculations.

The use of an effective chemical potential, which can take into account the interaction in the ideal gas equation of state, is also analyzed. It was found that under the conditions of applicability of the baryonic model, the effective chemical potential differs little from the potential obtained using the saddle point method. For the mesonic model, the use of an effective mesonic chemical potential is much less justified.

The research was carried out within the framework of the initiative scientific topic 0122U200549 (“National Technical University of Ukraine «Igor Sikorsky Kyiv Polytechnic Institute”).

\end{document}